\documentclass[times]{weauth}
\usepackage{moreverb}
\usepackage{subfigure}
\usepackage{graphicx}
\usepackage{dcolumn}
\usepackage{amssymb,amsmath}
\usepackage{epstopdf}
\usepackage{placeins}
\usepackage{citesort}
\usepackage{color}
\usepackage{epstopdf}

\def\volumeyear{2015}

\begin{document}

\runningheads{Stevens}{Dependence of optimal wind turbine spacing on wind farm length}
\articletype{\noindent "This is the peer reviewed version of the following article: \textbf{Stevens, R. J. A. M. (2016) Dependence of optimal wind turbine spacing on wind farm length. Wind Energ., 19: 651-663.}, which has been published in final form at \textbf{http://dx.doi.org/10.1002/we.1857}. This article may be used for non-commercial purposes in accordance with Wiley Terms and Conditions for Self-Archiving." \nobreak}
\title{Dependence of optimal wind turbine spacing on wind farm length}
\author{Richard J.A.M. Stevens}
\address{Department of Mechanical Engineering, Johns Hopkins University, Baltimore, Maryland 21218, USA\\
Department of Physics, Mesa+ Institute, and J.\ M.\ Burgers Centre for Fluid Dynamics, University of Twente, 7500 AE Enschede, The Netherlands}
\corraddr{r.j.a.m.stevens@jhu.edu}

\begin{abstract}
Recent large eddy simulations have led to improved parameterizations of the effective roughness height of wind farms. This effective roughness height can be used to predict the wind velocity at hub-height as function of the geometric mean of the spanwise and streamwise turbine spacings and the turbine loading factors. Meyers and Meneveau (Wind Energy 15, 305-317 (2012)) used these parameterizations to make predictions for the optimal wind turbine spacing in infinitely large wind farms. They found that for a realistic cost ratio between the turbines and the used land surface the optimal turbine spacing may be considerably larger than used in conventional wind farms. Here we extend this analysis by taking the length of the wind farm, i.e.\ the number of rows in the downstream direction, into account and show that the optimal turbine spacing strongly depends on the wind farm length. For small to moderately sized wind farms the model predictions are consistent with spacings found in operational wind farms. For much larger wind farms the extended optimal spacing found for infinite wind farms is confirmed. \\
\end{abstract}

\keywords{Wind-energy, turbine spacing, large eddy simulations, top-down model, turbine spacing}

\maketitle

\vspace{-0.4cm}
\section{Introduction} \label{Section_introduction}
Plans by the European \cite{eu2007} and American \cite{us2008} governments aim to increase the use of wind energy in the overall electricity production up to $20\%$ by 2030. To reach these targets larger onshore and offshore wind farms will be required \cite{arc14}. Due to the trend of increasing wind farm size there is interest in developing models that can predict the performance of large wind farms. Typically models that use algebraic expressions to model the wake and the superposition of the wakes or Reynolds Averaged Navier-Stokes (RANS) models are used \cite{bar11,san11}. More recently also large eddy simulations (LES) have become available \cite{jim07,jim08,cal10,cha10,mey10,cal11,mey12,por11,lee12,chu12,chu12b,wu13,ste13,ste14,ste14f}, although this method is still computationally very expensive. For reviews describing the different methods we refer to Refs.\ \cite{san11,meh14}.

In many cases it is challenging to extract the basic trends and the limiting behavior from these simulations. Therefore it is important to also have simple physics-based models available to improve our fundamental understanding of the basic trends and asymptotic behavior. The one-dimensional, single column type model proposed by Frandsen \cite{fra92,fra06} is such an approach. In this model the complex three-dimensional flow structure in the wind farm is simplified by using horizontal averaging. Subsequently an equation for the average streamwise velocity in the wind farm is obtained by vertically integrating the momentum equation. This approach allows one to include the effect of the wind farm on the flow in an atmospheric boundary layer. Describing a complex three-dimensional flow structure using horizontal averaging has important limitations, see for example the work by Lu and Port\'e-Agel \cite{lu11}. On the other hand the simplicity of such an approach is very useful to improve our understanding of the main physical mechanisms, which can be useful for the design and optimization of wind farms. 

In the Frandsen model \cite{fra92,fra06} it is assumed that there is a constant stress (momentum flux) above the turbine hub-height $z_\mathrm{h}$ with a characteristic friction velocity $u_{\mathrm{*hi}}$, while there is another constant stress region below the hub-height region with a characteristic friction velocity $u_{\mathrm{*lo}}$. The difference between the two momentum fluxes is given by the momentum extracted by the turbines, which is assumed to be concentrated in the hub-height plane. The Frandsen model \cite{fra92,fra06} gives a prediction for the effective roughness height $z_{\mathrm{0,hi}}$ of the wind farm, which can be used to describe the entire wind farm as a roughened surface with increased momentum flux and kinetic energy losses. The existence of these two logarithmic layers has been confirmed in LES \cite{cal10,wu13}. These LES results show that there is an intermediate wake region between the two logarithmic layers, which is not captured in the Frandsen model \cite{fra92,fra06}. Calaf {\it et al.} \cite{cal10} included the turbine wake layer into the model by noting that the turbine wakes create a region where the eddy mixing is increased, which is taken into account by an increased eddy viscosity. This improved model predicts the horizontal mean velocity distribution in the three layers, i.e.\ the bottom logarithmic layer, the turbine wake layer, and the top logarithmic region, see the sketch in figure \ref{figure2}. Comparisons with LES showed that the model provides a realistic prediction of the effective surface roughness created by infinitely large wind farms \cite{cal10}.

Meyers and Meneveau \cite{mey12} used the surface roughness model by Calaf {\it et al.} \cite{cal10} to predict the optimal turbine spacing in infinitely large wind farms as function of the turbine spacing while taking the ratio of the cost of the turbines and the cost of the land into account. They found an optimal spacing that is considerably larger, i.e.\ around $15$ turbine diameters, than used in current designs of wind farms \cite{mey12}. In order to explain this apparent discrepancy we investigate the influence of the wind farm length, i.e.\ in the number of rows in the downstream direction, on the optimal turbine spacing in this paper. For this purpose we use the notion of a developing internal boundary layer as introduced by Meneveau \cite{men12}. In section \ref{Section_model} we explain the physical model that is used to estimate the power production and compare the model predictions with measurements from Horns Rev, Nysted, \cite{bar11} and LES \cite{ste13,ste14f}. Here we follow the framework set out by Calaf {\it et al.} \cite{cal10} and Meneveau \cite{men12}. Based on the comparison between the model and LES some adjustments have been made which will be highlighted. In section \ref{Section_optimal} we investigate the optimal turbine spacing as function of the wind farm length and the turbine thrust coefficient $C_\mathrm{T}$ by taking the ratio of the cost of the turbines and the land into account \cite{mey12}, which is followed by the main conclusions in section \ref{Section_conclusion}. Before we introduce the model we will start in section \ref{section_LES} with a short description of the LES that are used for the comparison with the model.

\section{Large eddy simulation method} \label{section_LES}

In this paper we will compare the model with LES results for extended wind farms consisting of a regular array of wind turbines. The LES have been performed for wind turbines with a diameter $D$ and a hub-height $z_\mathrm{h}$ of $100$m. The surface roughness height in the LES was $z_\mathrm{0,lo}=0.1$m. Because a very large physical domain is considered the turbines are represented with an area averaged actuator disk approach \cite{jim07,cal10,yan13,ste14} using a turbine thrust coefficient $C_\mathrm{T}=0.75$, which is representative for turbines operating in region II \cite{joh04}. In the simulations wind farms with more than ten rows in the streamwise direction were considered in order to study the fully developed state, which is the regime in which the power output as function of the downstream position becomes constant. This was obtained by considering an LES domain of $8\pi \times \pi \times 2$ km ($L_\mathrm{x}\times L_\mathrm{y} \times L_\mathrm{z}$) in the streamwise, spanwise and vertical directions, respectively. The domain is discretized with up to $1536\times192\times384$ grid points using an uniform grid spacing in each direction. The subgrid scale stresses are modeled using the scale dependent Lagrangian dynamic subgrid scale model \cite{bou05}.

The inflow condition is generated with the concurrent precursor method described in Ref.\ \cite{ste14}. In this method the computational domain in the streamwise direction is divided in two sections. In the first section a neutral turbulent atmospheric boundary layer is simulated in a periodic domain using a pressure gradient forcing. Coriolis and thermal effects are not specifically included, an approach also used in previous studies such as \cite{cal10,wu11,lu11,cal14}. Each time step a part of the flow field from this simulation is used to provide the inflow condition for a second section in which the wind farm is placed. In the wind farm section, which is periodic due to the use of spectral methods in the horizontal direction, a long fringe region at the end of the computational section is used to make sure that there is a smooth transition from the flow formed behind the wind farm towards the applied inflow condition. The results are averaged over a period of about six hours \cite{ste14e}. Simulations for different combinations of the streamwise $s_\mathrm{x}$ and spanwise $s_\mathrm{y}$ turbine spacings have been performed. The considered dimensionless turbine spacings (in units of the rotor diameter $D$) are in the range $\sim[3.49,7.85]$. Relevant for the discussion presented here is the dimensionless geometric mean turbine spacing $s:=\sqrt{s_\mathrm{x} s_\mathrm{y}}$. Further details about the simulations can be found in \cite{ste14,ste14f}. 

The LES data described above were used by Stevens {\it et al.} \cite{ste14f} to show that for staggered wind farms the power output in the fully developed regime depends primarily on the geometric mean turbine spacing. This indicates that for large wind farms the geometric mean turbine spacing is an important design parameter.  A comparison \cite{ste14g} of the Calaf {\it et al.} \cite{cal10} model predictions with the LES data reveals that the performance of staggered wind farms is predicted particularly well, see also figure \ref{figure5}b. As for a given geometrical mean turbine spacing a staggered wind farm has a higher power output than an aligned wind farm the use of the Calaf {\it et al.} \cite{cal10} model approach is reasonable as and optimal wind farm design will be closer to a staggered arrangement than an aligned configuration. For aligned wind farms it has been shown that a higher power output is obtained for the same geometrical mean turbine spacing when the streamwise spacing is larger \cite{ste14f,yan12} and to make accurate predictions for the aligned configuration more advanced models are required. In the next section we will first introduce the model and compare with the results from LES at the end of the section.

\section{Model for wind farm optimization} \label{Section_model}

\begin{figure}
\begin{center}
\subfigure{\includegraphics[width=0.53\textwidth]{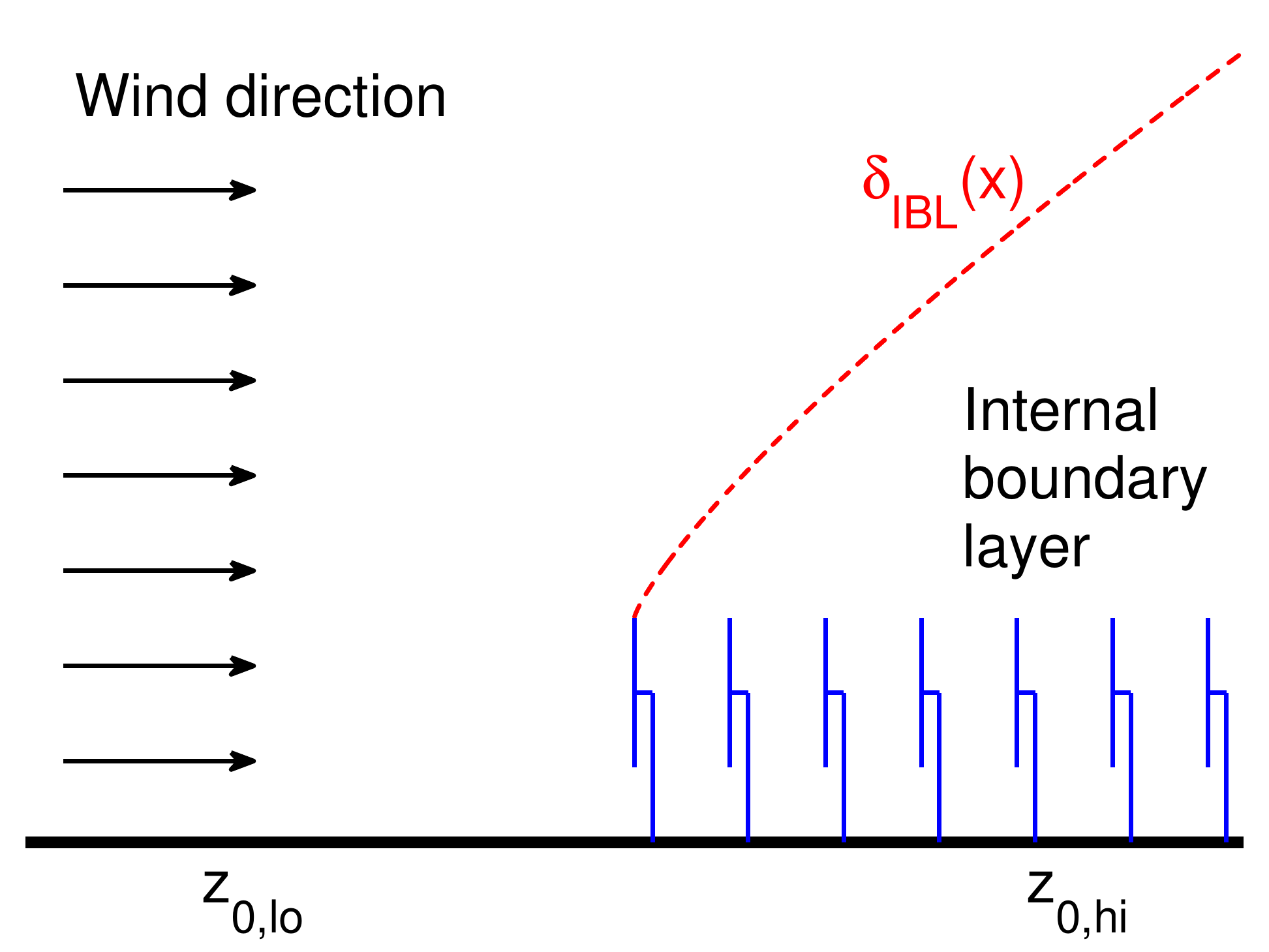}}
\caption{Sketch of the formation of an internal boundary layer at the start of a wind farm.}
\label{figure1}
\end{center}
\end{figure}

The streamwise velocity in the surface layer of the atmosphere is well described by the logarithmic law for the mean velocity \cite{pra25,kar30,mil38},
\begin{align}
\label{Eq_Inflow}
 \langle \overline{u_0} \rangle (z) = \frac{u_*}{\kappa} \ln \left( \frac{z}{z_{\mathrm{0,lo}} } \right) & ~~~~~~~~~~ \mathrm{for} & z_{\mathrm{0,lo}} & \leq z \leq H,
\end{align}
where $\kappa=0.4$ is the von K\'arm\'an constant, $z_{\mathrm{0,lo}}$ the roughness height of the ground, $u_*$ the friction velocity, and $H$ the height of the surface layer (the height of the computational domain in LES). As is shown in the sketch in figure \ref{figure1} an internal boundary layer begins to form at the leading edge of the wind farm and grows with increasing downstream position \cite{men12}. Above the internal boundary layer $\delta_{\mathrm{IBL}}(x)$ the boundary layer is undisturbed, while inside the internal boundary layer the flow structure changes due to the momentum that is extracted by the wind turbines. In order to predict the power production as function of the downstream position in the wind farm we therefore need to model the average flow structure inside the internal boundary layer and have an estimate for the internal boundary layer height. In this section we will first derive the horizontally averaged streamwise velocity profile in the internal boundary layer and compare this with data from LES. Subsequently, we will discuss the height development of the internal boundary layer based on LES data in section \ref{Section_IBL}. Finally, in section \ref{Section_comparison} we will compare the power production predicted by the model with data from the Horns Rev and Nysted wind farms \cite{bar11}, and LES \cite{ste13,ste14f}.

\subsection{Horizontally averaged velocity profiles} \label{Section_profiles}

\begin{figure}
\begin{center}
\subfigure{\includegraphics[width=0.68\textwidth]{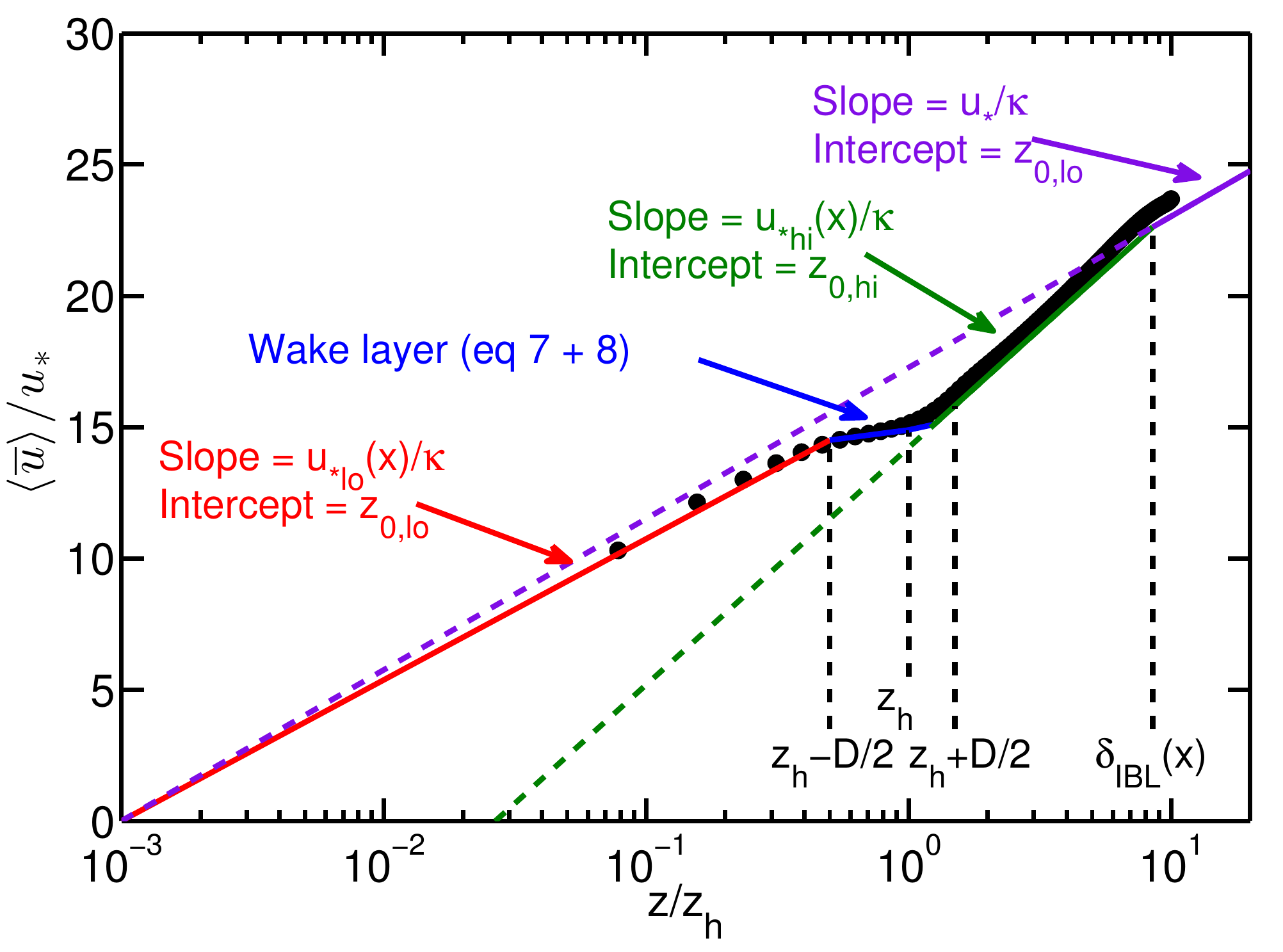}}
\caption{Sketch of the mean streamwise velocity profile compared with LES data for an infinite wind farm with $s_\mathrm{x}$=7.85 and $s_\mathrm{y}$=5.24. The lower logarithmic region eq.\ \eqref{Eq_profile_1} extends up to $z=z_\mathrm{h}-D/2$. The wake layer eq.\ \eqref{Eq_profile_2} and \eqref{Eq_profile_3} is observed between $z=z_\mathrm{h}-D/2$ and $z=z_\mathrm{h}+D/4$, and the logarithmic region above the turbine region eq.\ \eqref{Eq_profile_4} extends up to the height of the internal boundary layer $\delta_{\mathrm{IBL}}(x)$, above which the velocity profile is assumed to be undisturbed eq.\ \eqref{Eq_profile_5}. Note that the height of the internal boundary layer (eq.\ \eqref{Eq_model_delta_IBL}) and the slope of the logarithmic regions above and below the turbines depend on the downstream position in the wind farm according to eq.\ \eqref{Eq_model_continuity} to eq.\ \eqref{Eq_model_ushi}.}
\label{figure2}
\end{center}
\end{figure}
Both the model of  Frandsen \cite{fra92,fra06} and Calaf {\it et al.} \cite{cal10,men12} assume the presence of two constant stress layers, one above and one below the turbine region. The logarithmic region above the wind turbine array is characterized by an upper friction velocity $u_{\mathrm{*hi}}$, and the lower logarithmic region by a friction velocity $u_{\mathrm{*lo}}$. Using the horizontally averaged momentum balance to relate the difference between the two momentum layers leads to 
\begin{equation}
\label{Eq_momentum_balance}
u_{\mathrm{*hi}}^2= u_{\mathrm{*lo}}^2+ \frac{1}{2}c_{\mathrm{ft}} [\langle \overline{u} \rangle (z_\mathrm{h})]^2, 
\end{equation}
where $\langle \overline{u} \rangle$($z_\mathrm{h}$) is the horizontally and time averaged mean velocity at hub-height and $c_{\mathrm{ft}}= \pi C_\mathrm{T}/(4s_\mathrm{x}s_\mathrm{y})$. In the approach of Frandsen it was assumed that these two logarithmic regions meet at hub-height $z_\mathrm{h}$. However, in the more detailed model by Calaf {\it et al.} \cite{cal10,men12} also a wind turbine wake region, where increased mixing created by the wakes leads to flatter mean profiles, is incorporated. As is shown in  figure \ref{figure2} the layer below the turbines is a constant stress layer with a stress $u_{\mathrm{*lo}}^2$ and an eddy viscosity of $\kappa z u_{\mathrm{*lo}}$, which allows one to write an equation for the horizontally averaged velocity profile $({\kappa z u}_{\mathrm{*lo}}) {d \langle \overline{u} \rangle }/{dz}=u_{\mathrm{*lo}}^2$, which can be integrated from the ground to give:
\begin{align}
\label{Eq_profile_1}
\langle \overline{u} \rangle (z) = \frac{ u_{\mathrm{*lo}} }{k} \ln \left( \frac{z}{z_{\mathrm{0,lo}} } \right) & ~~~~~~~~~~~~~~\mathrm{for} &z_{\mathrm{0,lo}} &\leq z \leq z_\mathrm{h}-\frac{D}{2}. 
\end{align}
The profile in the logarithmic region above the wind turbines is another constant stress layer, now with a stress of $u_{\mathrm{*hi}}^2$ and with an eddy viscosity $\kappa z u_{\mathrm{*hi}}$, can be obtained in a similar way from $({\kappa zu}_{\mathrm{*hi}}) {d \langle \overline{u} \rangle }/{dz}=u_{\mathrm{*hi}}^2$, and reads
\begin{align}
\label{Eq_profile_4}
\langle \overline{u} \rangle (z) = \frac{ u_{\mathrm{*hi}}}{k} \ln \left( \frac{z}{z_{\mathrm{0,hi}} } \right) & ~~~~~~~~~~~~~~ \mathrm{for} &z_\mathrm{h}+\frac{D}{4} &\leq z \leq \delta_{\mathrm{IBL}}(x)
\end{align}
where $z_{\mathrm{0,hi}}$ indicates the effective roughness of the wind farm. Here, we note that the original derivation by Calaf {\it et al.} \cite{cal10} this layer was assumed to start at $z_\mathrm{h}+D/2$. A detailed look at the LES results shown in figure \ref{figure2} indicates that the upper logarithmic region seems to extend further downward. Therefore we have assumed here that this logarithmic layer starts at $z_\mathrm{h}+D/4$ instead of $z_\mathrm{h}+D/2$ to better represent this effect for this case. Outside the internal boundary layer the flow is assumed to be undisturbed, i.e.\
\begin{align}
\label{Eq_profile_5}
\langle \overline{u} \rangle (z) = \frac{ u_{*}}{k} \ln \left( \frac{z}{z_{\mathrm{0,lo}} } \right) & ~~~~~~~~~~~~~~\mathrm{for} &\delta_{\mathrm{IBL}}(x) &\leq z \leq H.
\end{align}
The horizontally averaged velocity profiles in the wake region can be obtained by assuming that the eddy viscosity is increased by an additional wake eddy viscosity $\nu_w$, which gives \cite{cal10}
\begin{align}
(k z u_* + \nu_w) \frac{d \langle \overline{u} \rangle}{dz}= u_*^2 \rightarrow (1 + \nu_w^*) \frac{d \langle \overline{u} \rangle }{d \ln (z / z_\mathrm{h})} = \frac{u_*}{k} & ~~~~~~~~~~~~~~\mathrm{for} & z_\mathrm{h}-{\frac{D}{2}} \leq z \leq z_\mathrm{h}+\frac{D}{4}.
\end{align}
where $\nu_w^*= \nu/(k u_*z) \approx \sqrt{ \frac{1}{2} c_{\mathrm{ft}} } \langle \overline{u} (z_\mathrm{h}) \rangle D/ (k u_* z_\mathrm{h})$ (see Calaf {\it et al.} \cite{cal10} for details). As $\nu_w^*$ depends on the horizontally averaged velocity at hub-height $\langle \overline{u} (z_\mathrm{h}) \rangle$ an iterative procedure is used to converge $\nu_w^*$ and $\langle \overline{u} (z_\mathrm{h}) \rangle$. Here we note that using $\nu_*=28 \sqrt{\frac{1}{2} c_{\mathrm{ft}}}$ \cite{cal10}, which is obtained by assuming $z_\mathrm{h}/z_{\mathrm{0,hi}}=1000$, is a good approximation. For $C_\mathrm{T}=0.75$ and $z_{\mathrm{0,lo}}$ this approximation is tested as function of the geometric mean turbine spacing and it is found that the difference in the obtained power output with respect to the actual iteration procedure is (at most) in the order of $1\%$. This approximation can thus be used for simplicity when needed. In the wake layer the friction velocity is assumed to be $u_{\mathrm{*lo}}$ for $z<z_\mathrm{h}$ and $u_{\mathrm{*hi}}$ for $z>z_\mathrm{h}$. Vertically integrating this wake layer, and matching the velocities at $z=z_\mathrm{h}-D/2$ and $z=z_\mathrm{h}+D/4$ gives
\begin{align}
\label{Eq_profile_2}
\langle \overline{u} \rangle (z)= \frac{ u_{\mathrm{*lo}}}{k} \ln \left[ \left( \frac{z}{z_\mathrm{h}} \right)^{\frac{1}{1+\nu_\mathrm{w}^*}} \left( \frac{z_\mathrm{h}}{z_{\mathrm{0,lo}} } \right) \left( 1 - \frac{D}{2z_\mathrm{h}} \right)^{\beta} \right] 	& ~~~~~~~~~~~~~~\mathrm{for}&z_\mathrm{h}-\frac{D}{2} &\leq z \leq z_\mathrm{h},
\end{align}
and
\begin{align}
\label{Eq_profile_3}
\langle \overline{u} \rangle (z) = \frac{ u_{\mathrm{*hi}}}{k} \ln \left[ \left( \frac{z}{z_\mathrm{h}} \right)^{\frac{1}{1+\nu_\mathrm{w}^*}} \left( \frac{z_\mathrm{h}}{z_{\mathrm{0,hi}}} \right) \left( 1 + \frac{D}{4z_\mathrm{h}} \right)^{\beta} \right] 	& ~~~~~~~~~~~~~~\mathrm{for} &z_\mathrm{h} 		 &\leq z \leq z_\mathrm{h}+\frac{D}{4}.
\end{align}
Here the exponent $\beta$ is defined as $\beta=\nu_w^*/(1+\nu_w^*)$ \cite{cal10}. Equation \eqref{Eq_profile_3} allows us to write the velocity at hub-height as 
\begin{align}
\label{Eq_model_uhubheight}
\langle \overline{u} \rangle (z_\mathrm{h}) = \frac{u_{\mathrm{*hi}}}{k} \ln \left[ \left( \frac{z_\mathrm{h}}{z_{\mathrm{0,hi}}} \right) \left(1 + \frac{D}{4z_\mathrm{h}} \right)^\beta \right].
\end{align}
At $z=z_\mathrm{h}$ there should be continuity between equation \eqref{Eq_profile_2} and \eqref{Eq_profile_3}. This relationship gives
\begin{equation}
\label{Eq_model_continuity}
\frac{u_{\mathrm{*hi}}}{u_{\mathrm{*lo}}}= \left( \ln\frac{z_\mathrm{h}}{z_{\mathrm{0,lo}}} + \beta \ln\left[1-\frac{D}{2z_\mathrm{h}}\right] \right) / \left( \ln \frac{z_\mathrm{h}}{z_{\mathrm{0,hi}}} + \beta \ln\left[1+\frac{D}{4z_\mathrm{h}}\right] \right) . 
\end{equation}
Substituting this relationship in the momentum balance (eq.\ \eqref{Eq_momentum_balance}) and replacing the mean velocity at hub-height one gets for the roughness height
\begin{equation}
\label{Eq_model_z0hi}
z_{\mathrm{0,h}}= z_\mathrm{h} \left(1+\frac{D}{4 z_\mathrm{h}}\right)^{\beta} \\
\exp \left(- \left[ \frac{c_{\mathrm{ft}}}{2k^2} + \left( \ln \left[ \frac{z_\mathrm{h}}{z_{\mathrm{0,lo}}} \left( 1 -\frac{D}{2z_\mathrm{h}}\right)^{\beta}	\right] \right)^{-2}		\right]^{-1/2} 				\right).
\end{equation}
Note that this expression is slightly different than the one obtained by Calaf {\it et al.} \cite{cal10} and Meneveau \cite{men12}, because the logarithmic layer formed above the turbines is assumed to start at $z_\mathrm{h}+D/4$ instead of $z_\mathrm{h}+D/2$ based on the comparison with LES data shown in figure \ref{figure2}.

\subsection{Internal boundary layer height} \label{Section_IBL}

\begin{figure}
\begin{center}
\subfigure{\includegraphics[width=0.47\textwidth]{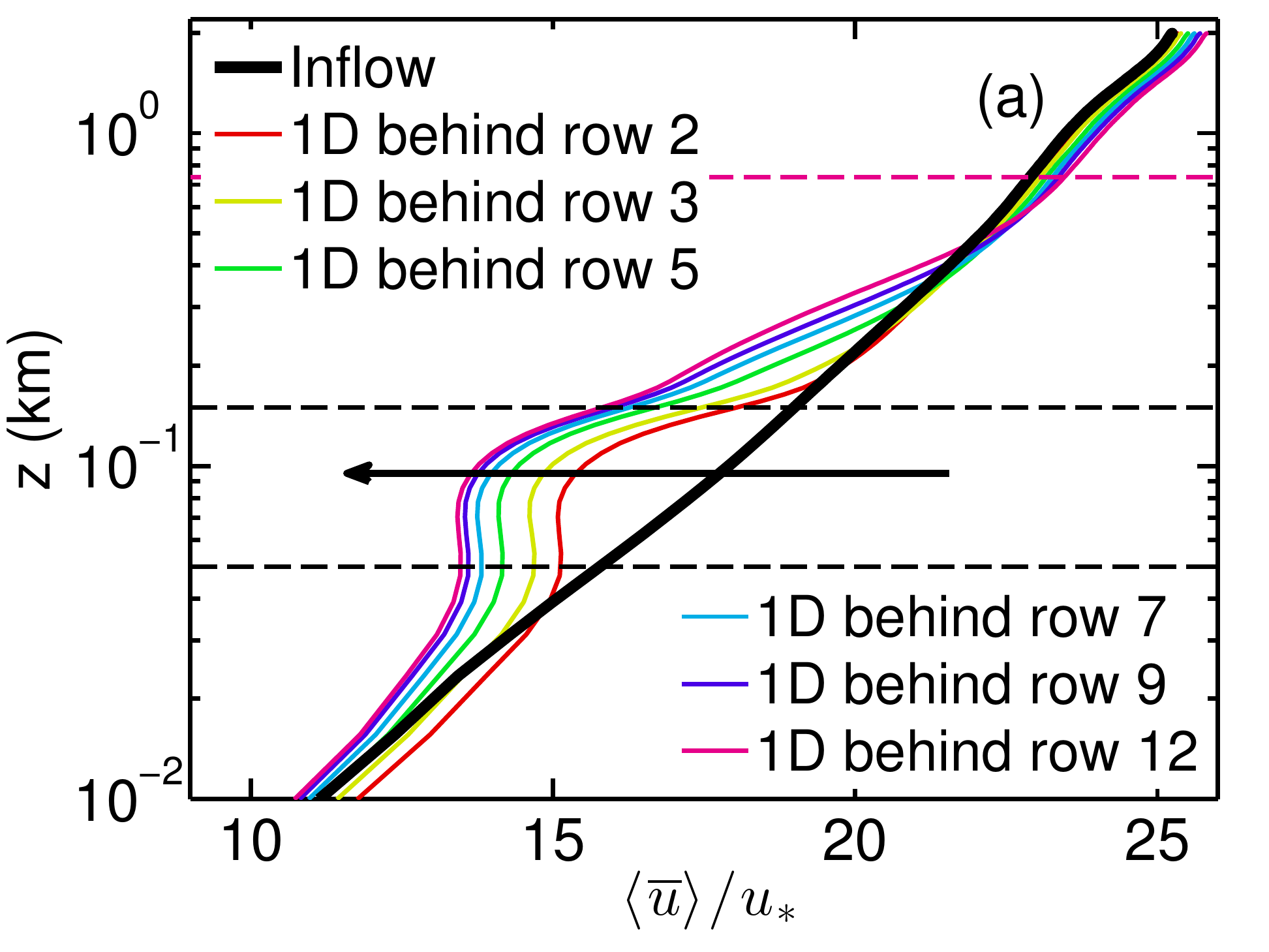}}
\subfigure{\includegraphics[width=0.47\textwidth]{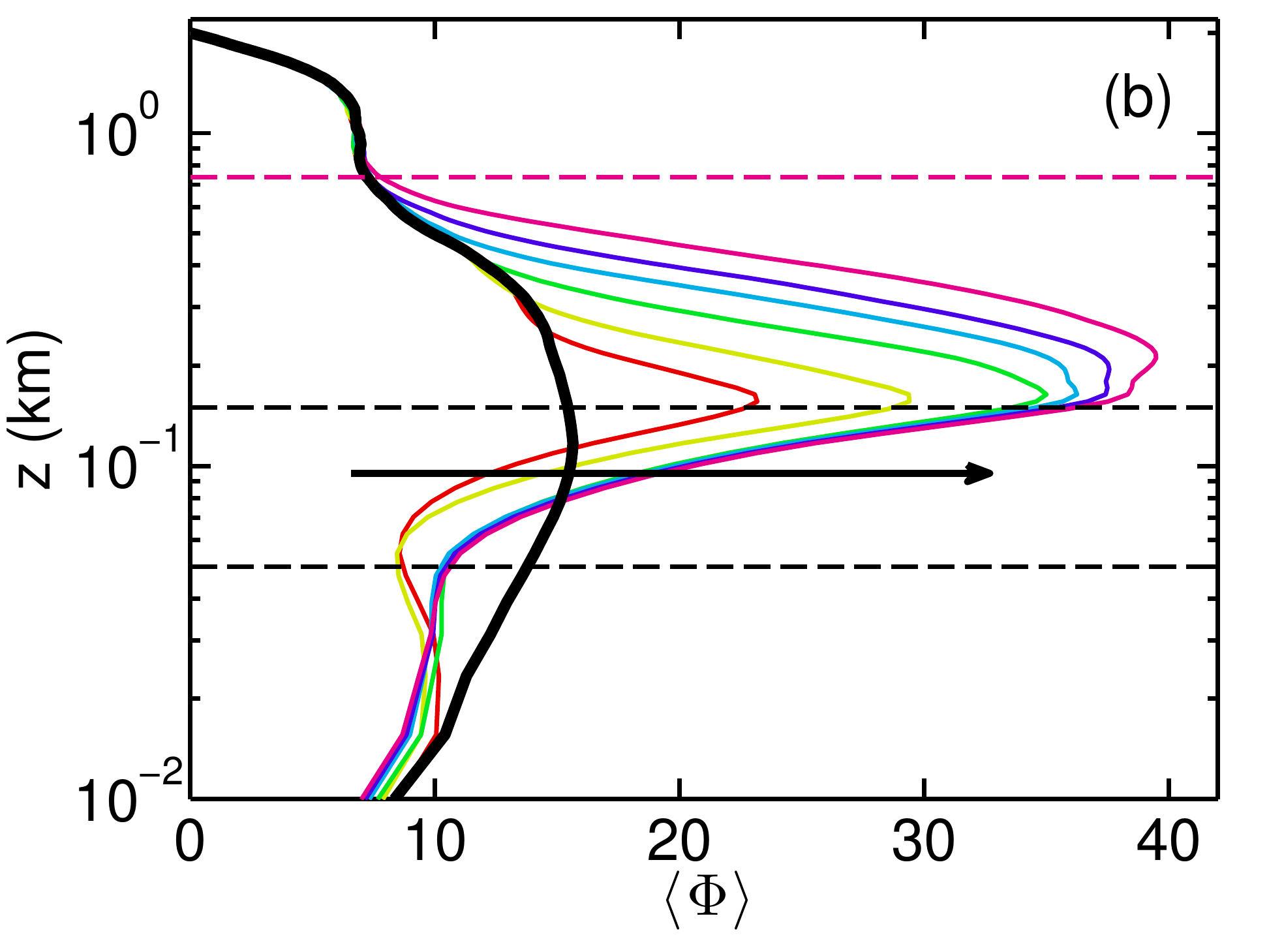}}
\caption{(a) The spanwise averaged streamwise velocity at 1D behind different turbine rows and (b) the corresponding vertical kinetic energy flux $\langle \Phi \rangle=\langle \overline{u^\prime w^\prime} \overline{u}  \rangle$ profiles as function of height for a staggered wind farm with $s_\mathrm{x}=7.85$ and $s_\mathrm{y}=5.24$. The dashed black horizontal lines indicate the vertical extension of the wind turbines and the dashed magenta horizontal line indicates $\delta_{\mathrm{IBL}}$ for the profile behind the $12^{\mathrm{th}}$ turbine-row. In both panels the inflow profile is shown by the thick black line.}
\label{figure3}
\end{center}
\end{figure}

\begin{figure}
\begin{center}
\subfigure{\includegraphics[width=0.47\textwidth]{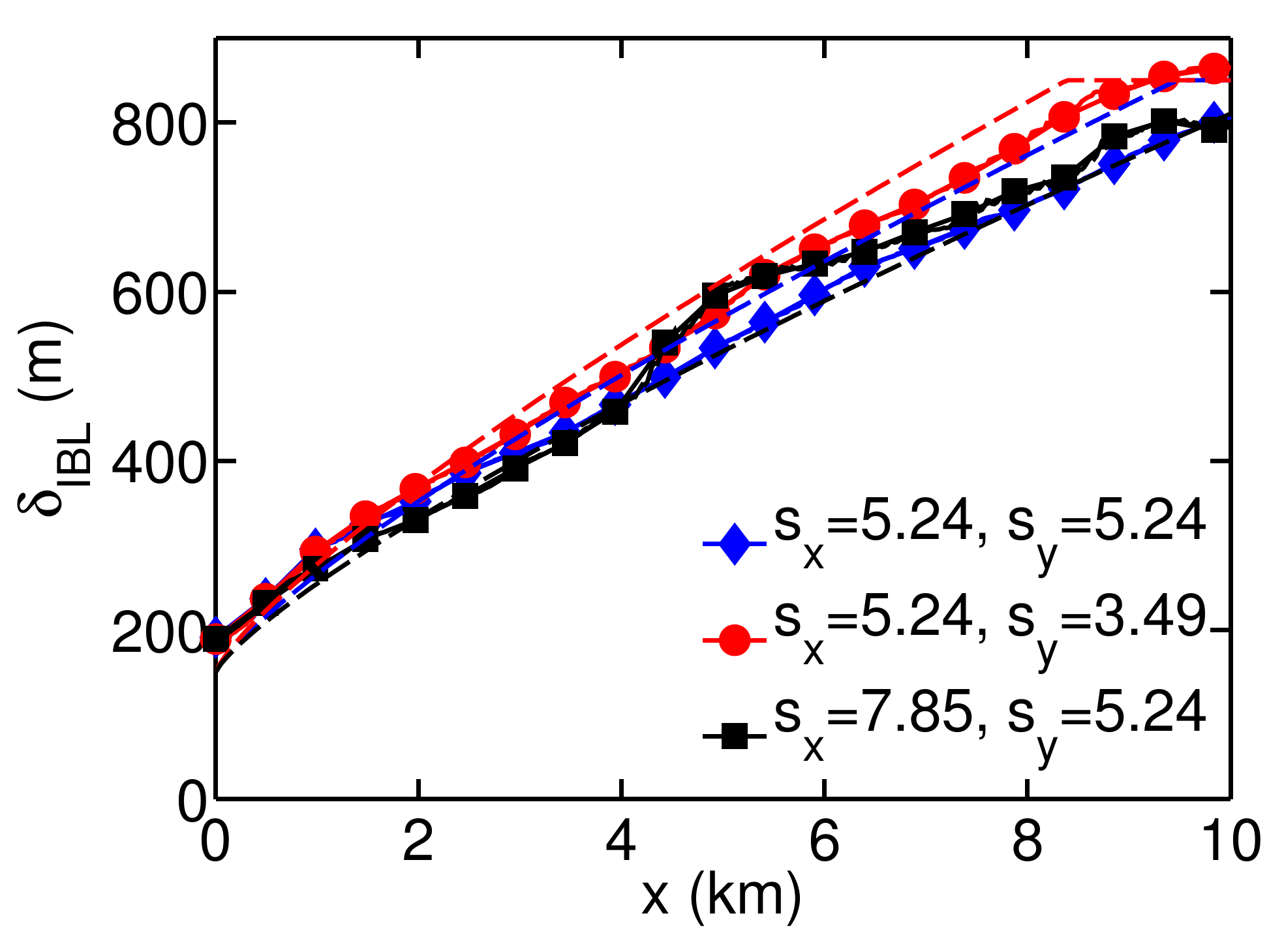}}
\caption{Internal boundary layer thickness obtained from the vertical kinetic energy profiles, see figure \ref{figure3}b, for three different staggered wind farms. The dashed lines indicate the development of the internal boundary layer thickness according to eq.\ \eqref{Eq_model_delta_IBL}.}
\label{figure4}
\end{center}
\end{figure}

In the previous section we derived the horizontally averaged streamwise velocity profile in the wind farm boundary layer. This velocity profile is only valid within the internal boundary layer. As indicated by Meneveau \cite{men12} the average velocity at hub-height and therefore the predicted power production depends on the height of the internal boundary layer. To find the mean velocity at hub-height we assume that the mean velocity from the logarithmic layer above the turbines vanishes at $z=z_{\mathrm{0,hi}}$ and that it merges with the unperturbed velocity at the internal boundary layer height $z=\delta_{\mathrm{IBL}}(x)$, see figure \ref{figure2}. This approach \cite{men12} gives
\begin{equation}
\label{Eq_model_ushi}
u_{\mathrm{*hi}}(x)= u_* \frac{\ln \left(\delta_{\mathrm{IBL}}(x)/z_{\mathrm{0,lo}} \right)} {\ln \left( \delta_{\mathrm{IBL}}(x)/z_{\mathrm{0,hi}} \right)}.
\end{equation}
Apart from the internal boundary layer height all coefficients of the model are known.

Classically the height of the internal boundary layer at a downstream location $x$ is parametrized as \cite{wil14}
\begin{equation}
\label{Eq_model_delta_IBL}
\frac{\delta_{\mathrm{IBL}}(x)}{z_{\mathrm{0,hi}}} = \frac{\delta_{\mathrm{IBL}}(0)}{z_{\mathrm{0,hi}}} + C_\mathrm{1} \left ( \frac{x}{z_{\mathrm{0,hi}}} \right) ^{4/5}.
\end{equation}
In order to get an estimation for the internal boundary layer height we look at data obtained from LES. Figure \ref{figure3} shows the spanwise averaged streamwise velocity at different downstream distances for a wind farm with a streamwise and spanwise spacing of 7.85D and 5.24D, respectively. Figure \ref{figure3}b shows the corresponding vertical kinetic energy flux $\langle \Phi \rangle = \langle \overline{u^\prime w^\prime} \overline{u} \rangle$ profiles.  The mean wind speed vanishes close to the ground, while the turbulent flux vanishes at the top of the model domain. Therefore the vertical kinetic energy flux vanishes at the top and the bottom of the domain, while it is non-zero in between. We determine the internal boundary layer thickness as the height at which the vertical kinetic energy flux becomes equal to the vertical kinetic energy flux of the inflow profile within $99\%$ and requiring that $\delta_\mathrm{IBL}>z_\mathrm{h}+D/2$. The development of the internal boundary layer thickness for different wind-farm densities is shown in figure \ref{figure4}. Up to an internal boundary layer thickness of 850 m the results are approximated well with $C_\mathrm{1}=1/3$ and the initial boundary layer thickness is set to $\delta_{\mathrm{IBL}}(0)=z_\mathrm{h}+D/2$. This result is close to the $C_\mathrm{1}=0.28$ value found by \cite{wil14}. The maximum internal boundary layer thickness is set to $850$ meters.

\subsection{Comparison with data} \label{Section_comparison}

To predict the power as function of the downstream position we calculate the power output of the different rows by comparing the inflow hub-height velocity with the velocity at hub-height obtained from the model (eq.\ \eqref{Eq_model_uhubheight}). Thus the normalized power output for different turbine rows can be calculated as
\begin{equation}
\label{Eq_model_prediction}
\frac{P}{P_\mathrm{0}} = \left[ \frac{u_{\mathrm{*hi}}(x)}{\kappa} \ln \left[ \left( \frac{z_\mathrm{h}}{z_{\mathrm{0,hi}}} \right) \left(1 + \frac{D}{4z_\mathrm{h}} \right)^\beta \right]~~ / ~~ \frac{u_*}{k} \ln{ \left( \frac{z_\mathrm{h}}{z_{\mathrm{0,lo}}}\right)} \right]^3.
\end{equation}
Here $P_\mathrm{0}$ is the power output for the first turbine row. The dependence on the downstream position in the wind farm is captured by the prediction for the average velocity at hub-height. As is shown in equation \eqref{Eq_model_prediction} this average velocity at hub-height depends on the downstream location in the wind farm. Equation \eqref{Eq_model_ushi} gives that the dependence of $u_\mathrm{*hi}$ on $x$ is determined by the development of the internal boundary layer thickness as well as the roughness height of the wind farm. In the previous section we have seen that the parametrized internal boundary layer thickness is in good agreement with the trend obtained from LES. Here we also note that the approximate $\nu_*=28 \sqrt{\frac{1}{2} c_{\mathrm{ft}}}$ assumes the wind farm roughness height is constant throughout the wind farm while the iterative procedure results in a small variation in the roughness height of the wind farm as function of the downstream position consistent with the $1\%$ difference in the predicted power output.

\begin{figure}
\begin{center}
\subfigure{\includegraphics[width=0.47\textwidth]{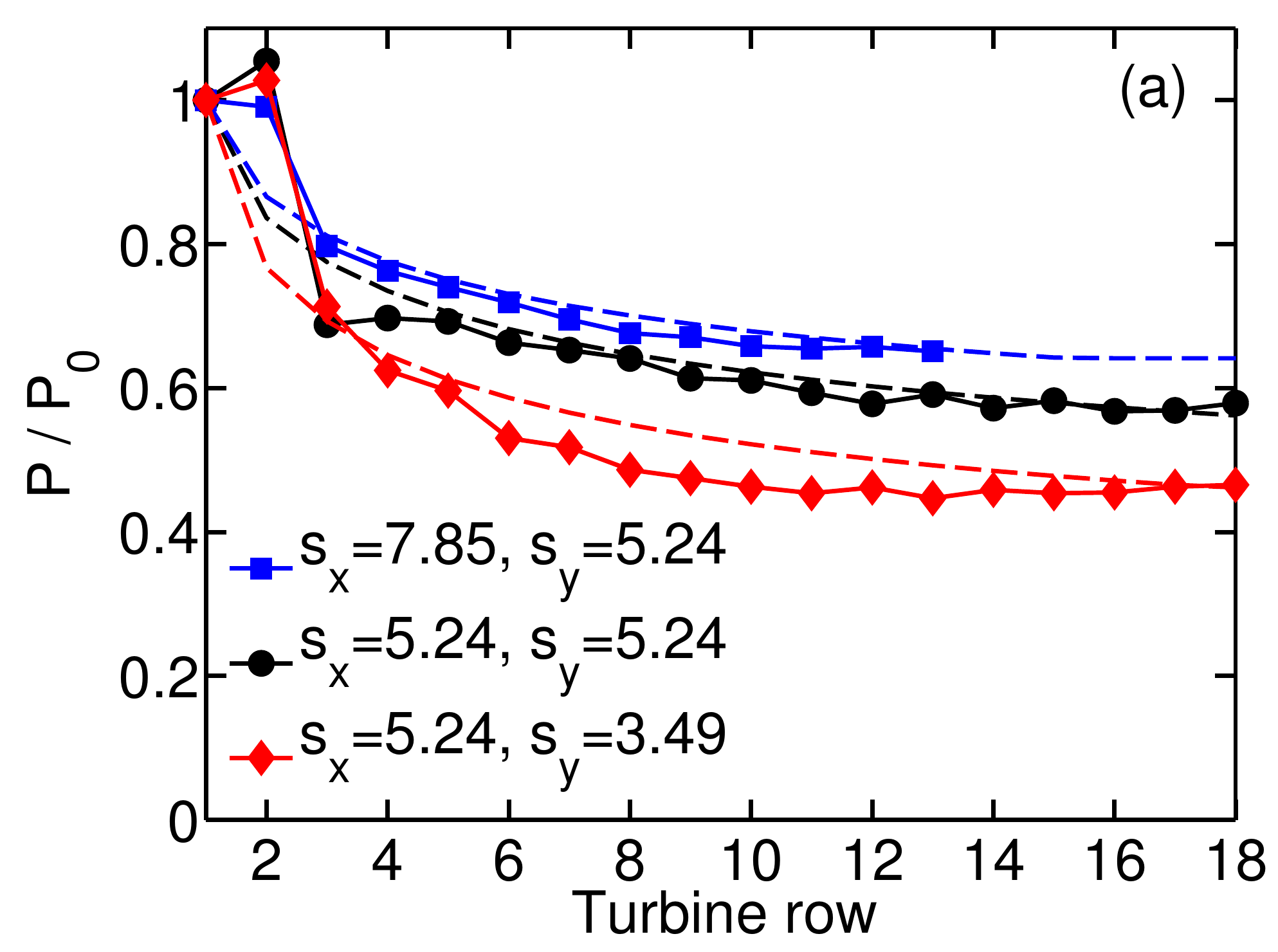}}
\subfigure{\includegraphics[width=0.47\textwidth]{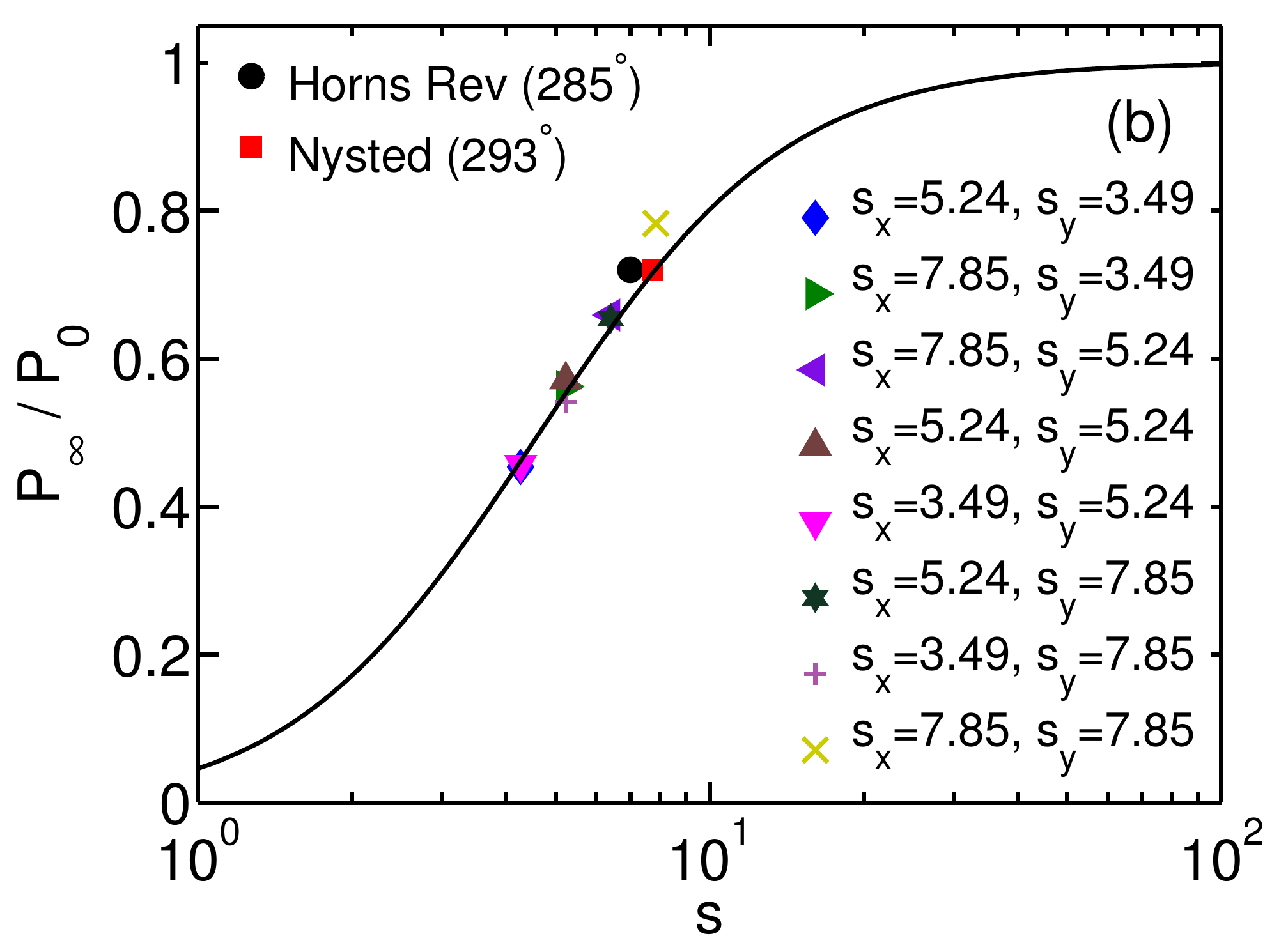}}
\caption{(a) The normalized power output $P/P_\mathrm{0}$ as function of the downstream position for staggered wind farms with different streamwise $s_\mathrm{x}$ and spanwise $s_\mathrm{y}$ spacing obtained from LES \cite{ste13,ste14f}. The streamwise and spanwise spacings are normalized with the rotor diameter. The dashed lines indicate the model prediction. (b) Comparison of the normalized power output in the fully developed regime predicted with the model  (eq.\ \eqref{Eq_model_prediction}, solid black line) with results from LES and observations in Horns Rev (285$^\circ$) and Nysted (293$^\circ$) \cite{bar11,bar09c}, see also the appendix.}
\label{figure5}
\end{center}
\end{figure}

Before we compare the model with data it is important to realize that the model considers horizontal averaging. This means the model gradually extracts energy from the flow with increasing distance in the wind farm. Therefore the model cannot predict differences due to the relative placement of the turbines. As the model gradually extracts energy from the flow it is closer to a staggered wind farm configuration than an aligned configuration, because in staggered wind farms the energy extraction is more homogeneously distributed in the horizontal plane than in aligned wind farms. To be specific; a pronounced difference between the staggered and the aligned configuration is that in an aligned wind farm the formation of high velocity wind channels in between the turbines is much stronger than for a staggered configuration \cite{ste14b}. Therefore figure \ref{figure5}a compares the model predictions for the power production with LES results of staggered wind farms with different streamwise and spanwise spacing \cite{ste13,ste14f}. The overall trend is predicted well by the model. One can notice that the model does not describe the small increase in the power production that can sometimes be seen at the second row. We note that the observation of this small increase at the second row is consistent with experimental results and is due to the flow acceleration in between turbines placed on the first row \cite{mye12,mct13,mct14}. For the highest turbine density we see that the model slightly overpredicts the power output for some of the intermediate rows, while the power output in the fully developed regime is predicted well. Figure \ref{figure5}b shows that the power production in the fully developed regime obtained from the model compares well with our LES results obtained for staggered wind farms \cite{ste13,ste14f}. In addition we compare this with measurements from Horns Rev and Nysted \cite{bar11}. For comparison we select the closest available wind direction to the staggered configuration, i.e.\ $285^{\circ}$ for Horns Rev and $293^{\circ}$ for Nysted, see the appendix. The figure shows that the LES and model results presented here are consistent with the field measurement data.

\section{Optimal turbine spacing} \label{Section_optimal}
In this chapter we will explain how the Calaf {\it et al.}  \cite{cal10} model approach introduced above can be used to predict the optimal turbine spacing as function of the wind farm length while taking the cost ratio of the turbines and the land into account (section \ref{section31}). Subsequently the effect of the wind farm length (section \ref{section32}) and the turbine thrust coefficient (section \ref{section33}) on the obtained optimal spacing will be presented.

\subsection{Incorporation of turbine and land cost} \label{section31}

For the design of wind farms it is important to know the cost of the energy. In the literature the cost for wind farms is estimated with varying complexity \cite{ema10,mar08,gra05,att11,hea11,mey12}. Estimates vary from only taking a fixed cost per turbine into account up to the complete modeling of the maintenance and electrical grid connection costs. In this work we use the relatively simple approach that captures dominant effects. Following the work of Meyers and Meneveau \cite{mey12} we define the cost as 
\begin{equation}
\label{Eq_cost_definition}
\mathrm{Cost} = \mathrm{Cost}_{\mathrm{land}} [\$/ \mathrm{m^2} ] \times S + \mathrm{Cost}_{\mathrm{turb}} [\$].
\end{equation}
Here $S$ is the average land surface area per turbine as defined before. This allows us to express the cost in a dimensionless ratio $\alpha$ defined as
\begin{equation}
\label{Eq_alpha_definition}
\alpha= \frac{\mathrm{Cost}_{\mathrm{turb}} / (\frac{1}{4} \pi D^2 ) }{\mathrm{Cost}_{\mathrm{land}}}.
\end{equation}
Subsequently, this dimensionless cost ratio $\alpha$ is used to define the normalized power per unit cost as

\begin{figure}
\begin{center}
\subfigure{\includegraphics[width=0.47\textwidth]{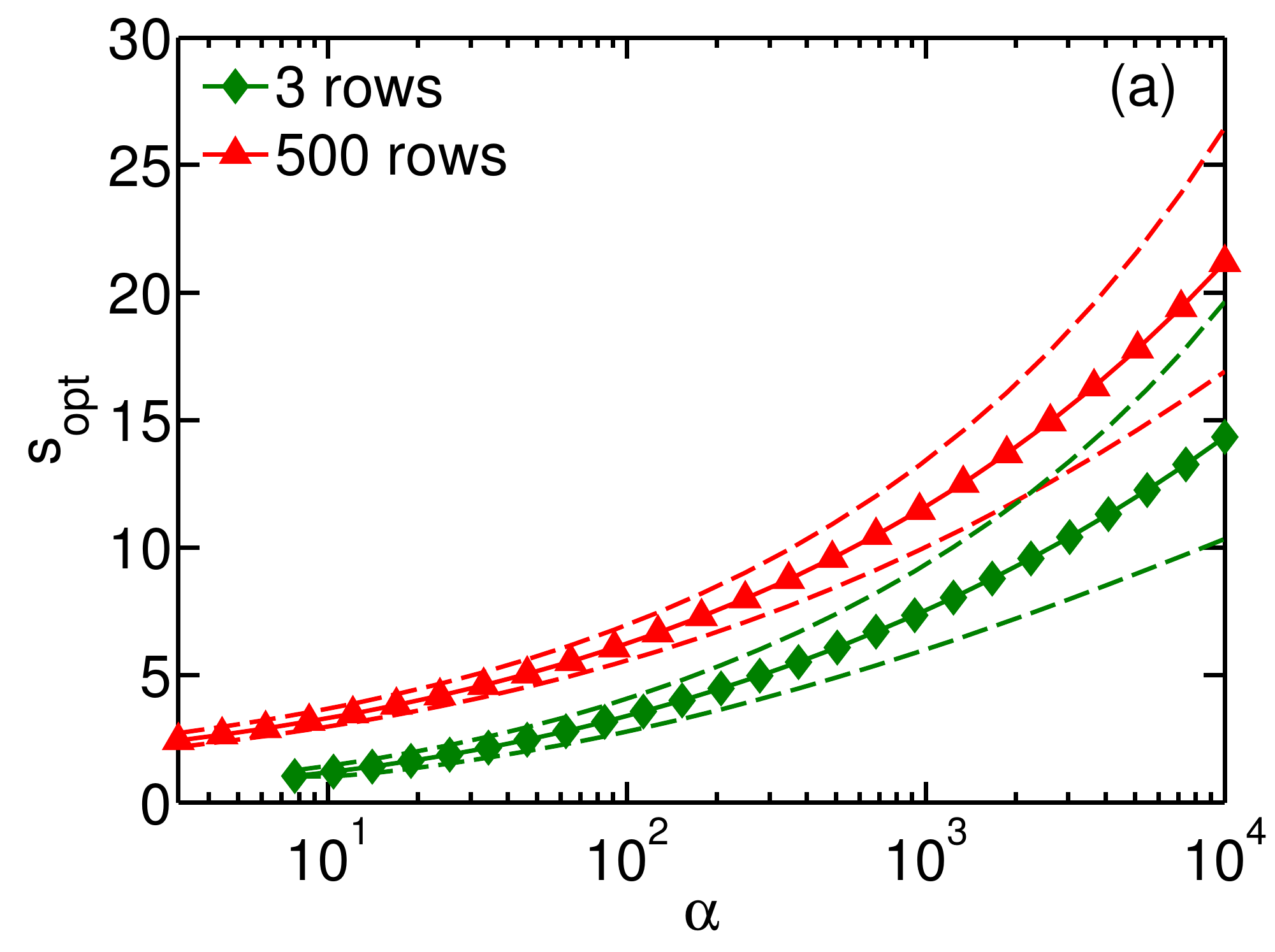}}
\subfigure{\includegraphics[width=0.47\textwidth]{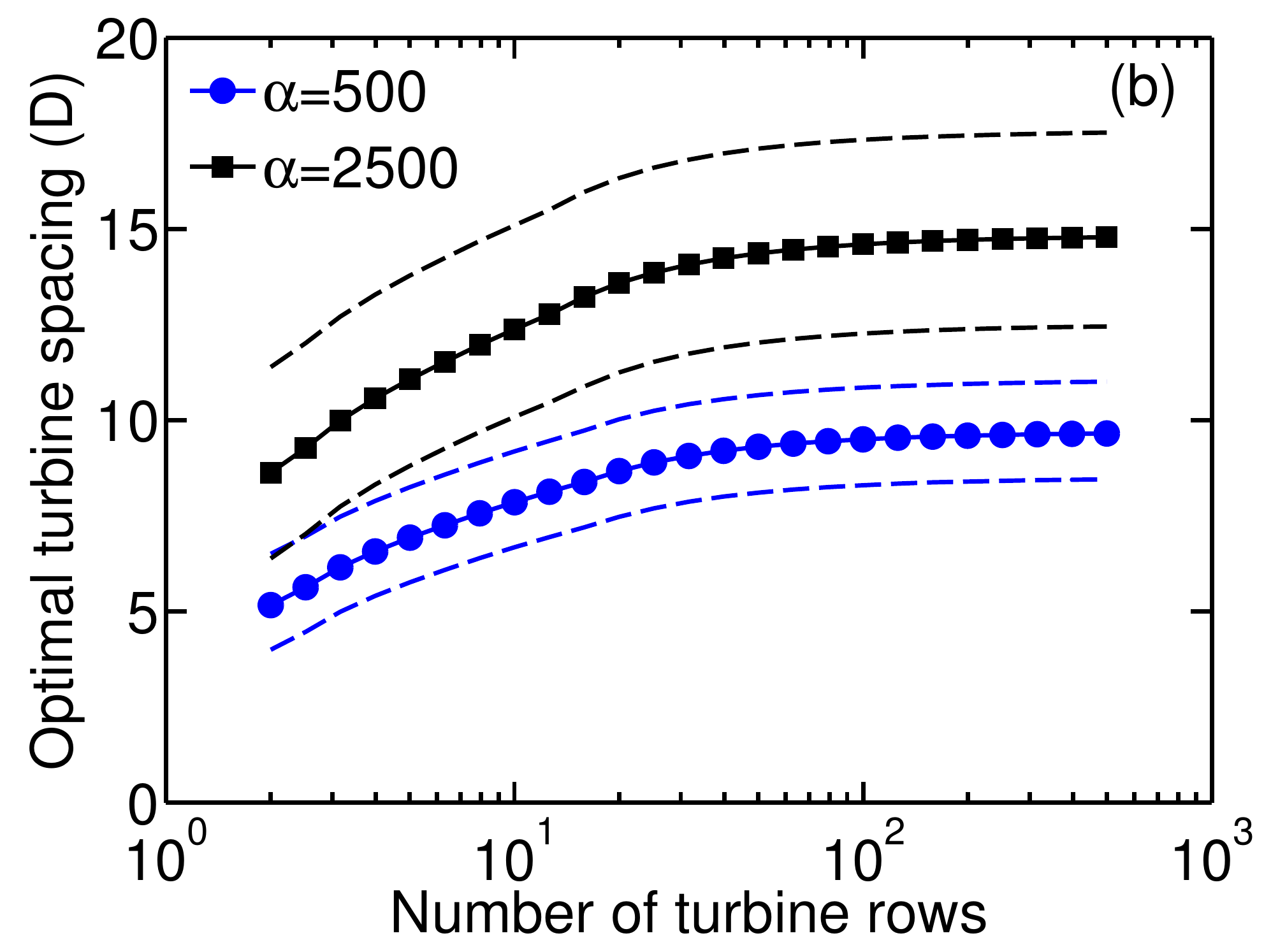}}
\caption{(a) Comparison of the optimal wind turbine spacing as function of the cost ratio $\alpha$ for a wind farm with $3$ rows and a wind farm with 500 rows. (b) The optimal wind turbine spacing as function of the number of wind turbine rows for $\alpha=500$ and $\alpha=2500$. The colored solid lines with symbols indicate the optimal value that is obtained. The corresponding dashed colored lines indicate the upper and lower border of the region where $P^*\gtrsim 99\% P^{*}_{\mathrm{max}}$.}
\label{figure6}
\end{center}
\end{figure}
\begin{equation}
\label{Eq_optim_definition}
P^{*}=\frac{P_{\mathrm{avg}}}{s_\mathrm{x}s_\mathrm{y}} \frac{ \mathrm{Cost}_{\mathrm{land}}}{\mathrm{Cost}_{\mathrm{turb}}/S + \mathrm{Cost}_{\mathrm{land}}} = \frac{P_{\mathrm{avg}}}{s_\mathrm{x}s_\mathrm{y}} \frac{4s^2/\pi}{\alpha + 4s^2/\pi},
\end{equation}
where $P_{\mathrm{avg}}$ is the average turbine power output in the wind farm normalized by the power output of a free standing wind turbine. Equation \eqref{Eq_optim_definition} allows us to find the normalized power per unit cost as function of the average turbine spacing as well as the wind farm length by determining the average power output for all cases using the model described above. The dimensionless cost ratio that should be used can depend on many factors such as the location of the wind farm and the type of turbine that is used. Meyers and Meneveau \cite{mey12} estimated that the dimensionless cost ratio $\alpha$ for a wind farm in Texas should be around 2000-3500. Their estimates were based on an average turbine cost of $\$ 3.5\times10^6$ for a wind turbine with a diameter of $70$m and a cost of $\$0.25\mathrm{m}^{-2}$ to $\$0.4\mathrm{m}^{-2}$ for the land. As the turbine and land cost can vary the results will also be discussed as function of $\alpha$.

\subsection{Influence of the wind farm length and the dimensionless cost ratio on optimal turbine spacing} \label{section32}

Figure \ref{figure6}a shows the optimal wind turbine spacing for a wind farm of three rows and for a wind farm with 500 rows as function of the dimensionless cost ratio $\alpha$. The solid lines with symbols show the optimal spacing, while the corresponding dashed colored lines indicate the corresponding $99\% P^{*}_{\mathrm{max}}$ lines. The figure reveals that the optimal spacing becomes larger when the dimensionless cost ratio is increased and that the normalized power per unit cost varies relatively slowly as function of the geometric mean turbine spacing. The latter results in the fact that the cost optimum is not very sharp as is indicated in the figure. In addition, the figure reveals that the optimal spacing depends significantly on the wind farm length. This is shown more clearly in figure \ref{figure6}b, which shows the optimal turbine spacing as function of wind farm length for two dimensionless cost ratios. The reason is that with increasing wind farm length a higher percentage of the wind turbines experience wake effects. Thus for longer wind farms there is a larger benefit of placing the turbines further apart and this leads to a larger optimal spacing. Here it is important to note that when the wind farm length is taken into account the predicted optimal spacing is in agreement with values found in operational wind farms.

Based on the work by Denholm {\it et al.} \cite{den09} and information available on the internet (www.windpowerengineering.com, www.thewindpower.net) the spacing for several large wind farms in Texas has namely be determined to be in the range of $8$ to $12$ turbine diameters, i.e.\ consistent with the results found for smaller and intermediate wind farms. Some specific examples are the Roscoe wind farm with 627 wind turbines spread out over an area of almost 100.000 acres. As the average turbine diameter in that wind farm is approximately $68$ meters this corresponds to an average geometric mean turbine spacing of $\approx9.8$ turbine diameters. Two other examples are the Horse Hollow and Penascal wind farms. These wind farms contain respectively $421$ and $168$ wind turbines spread out over an area of respectively 47.000 and 30.000 acres. Using the average turbine diameter of $\approx82$ meters in Horse Hollow and $\approx92$ meters in Penascal this leads to average geometric turbine spacings of $8.2$D (Horse Hollow) and $9.2$D (Penascal).

The analysis also shows that the effect of the wind farm length on the optimal spacing is gradual and that, for some purposes, the infinite wind farm case is only approached when the number of turbine rows is around $30$. The turbine spacing found for a large number of rows is consistent with the values reported by Meyers and Meneveau \cite{mey12} for the infinite case. This analysis thus explains the apparent discrepancy of the infinite wind farm results with values used in actual wind farms. 

\begin{figure}
\begin{center}
\subfigure{\includegraphics[width=0.47\textwidth]{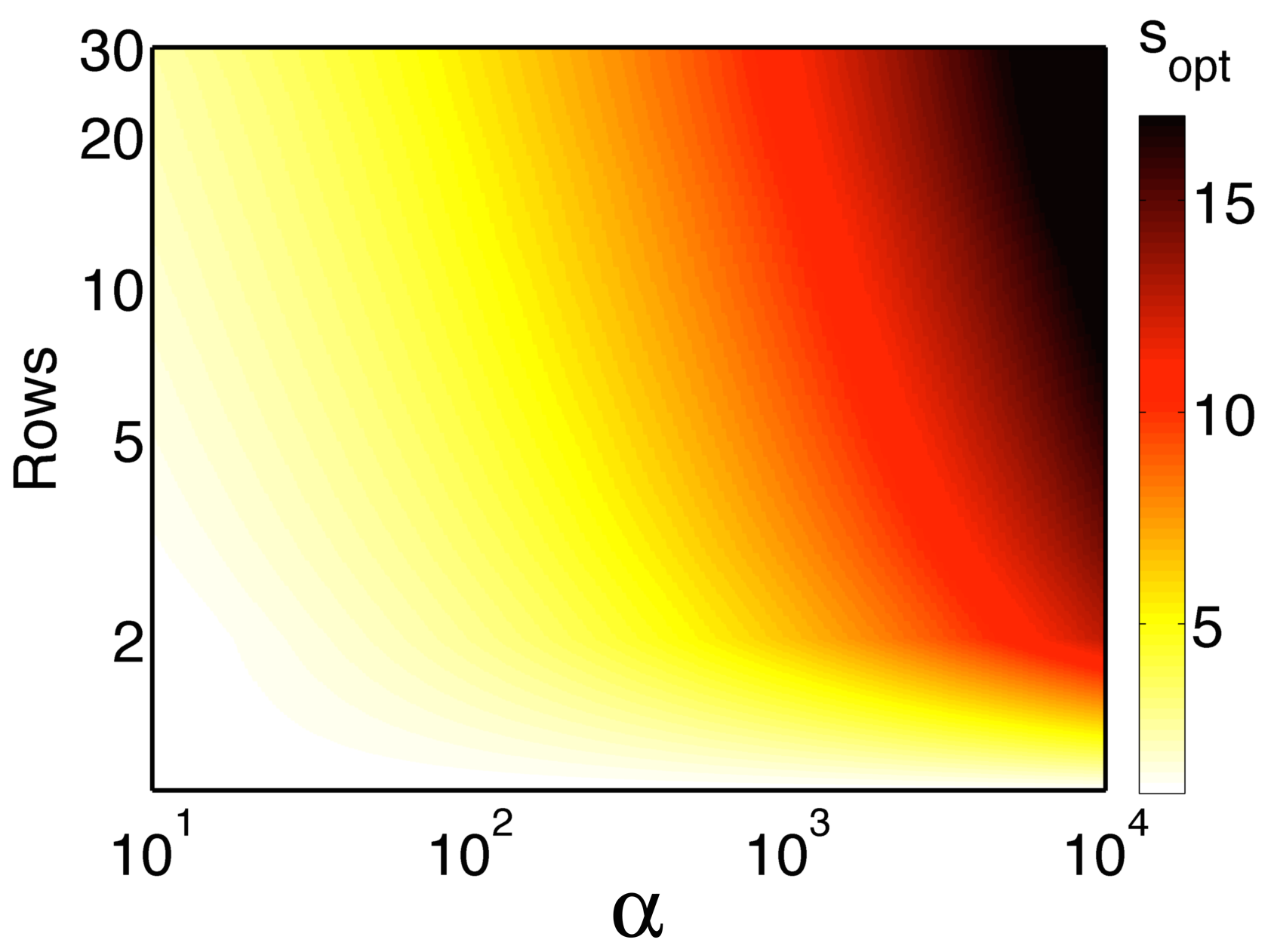}}
\caption{Optimal geometrical mean turbine spacing as function of the number of turbine rows and the dimensionless cost parameter $\alpha$ for $C_\mathrm{T}=0.75$.}
\label{figure7}
\end{center}
\end{figure}

\subsection{Influence of the turbine thrust coefficient on optimal turbine spacing} \label{section33}

Figure \ref{figure7} shows the optimal turbine spacing obtained from the model as function of the number of downstream turbine rows and the dimensionless cost parameter $\alpha$. In agreement with figure 6 it reveals that a small inter turbine distance should be used when turbines are cheap compared to the land (low $\alpha$). Note that $1$ turbine row corresponds to a line of turbines placed next to each-other. In this configuration the turbines do not experience any wake effects and therefore placing the turbines close together is beneficial as is shown in the figure. Wake effects become important for longer wind farms, in which more turbines are experiencing wake effects. The significant loss of performance when turbines are placed relatively close together leads to larger optimal turbine spacings when maximizing the normalized power per unit cost. This effect is most pronounced in the upper right corner of figure \ref{figure7}, i.e.\ when turbines are very expensive compared to the land and when the wind farm is very large. For lower $\alpha$ the effect of the wind farm length is less important as placing cheap turbines relatively close together is beneficial. 

\begin{figure}
\begin{center}
\subfigure{\includegraphics[width=0.47\textwidth]{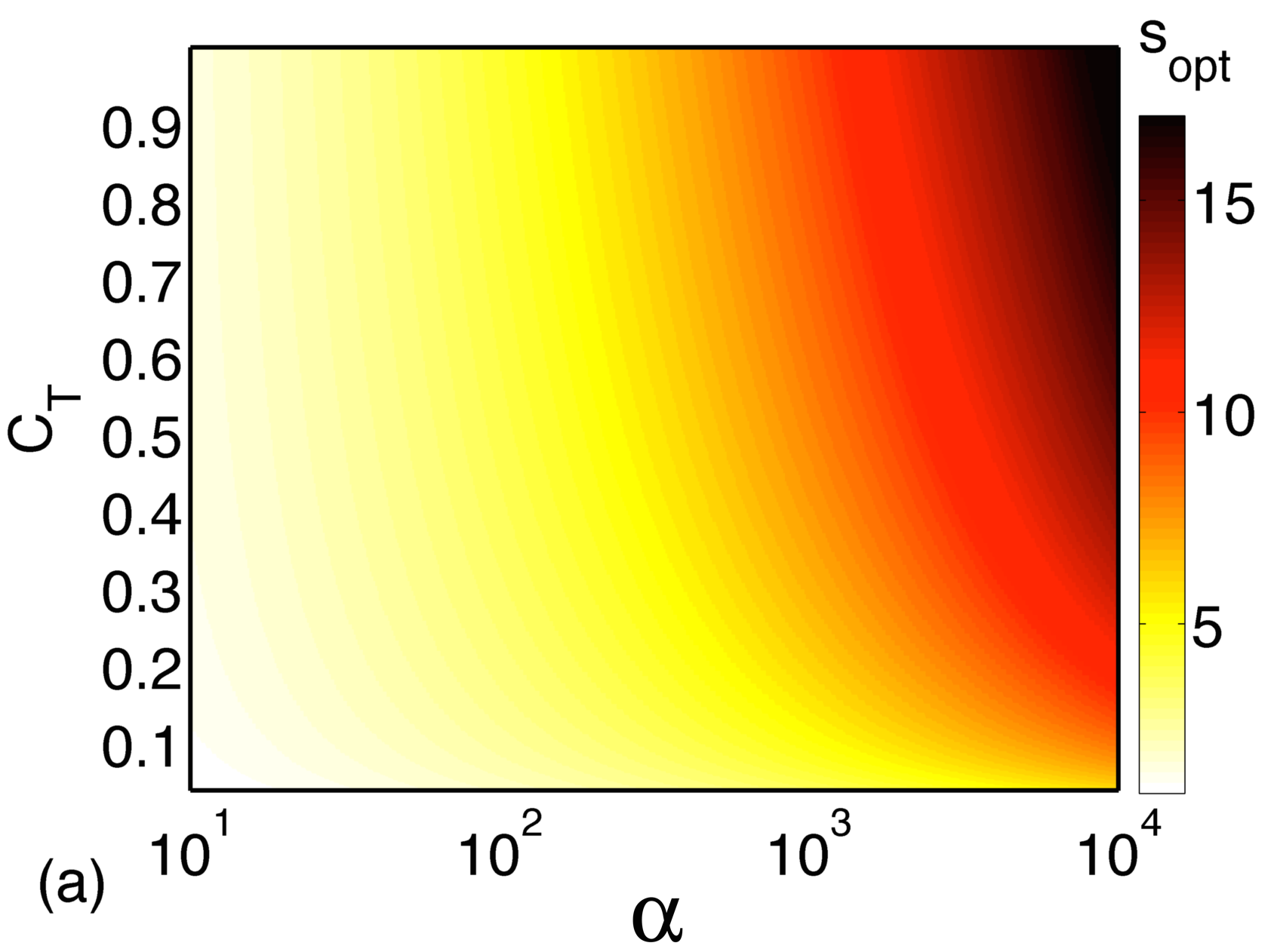}}
\subfigure{\includegraphics[width=0.47\textwidth]{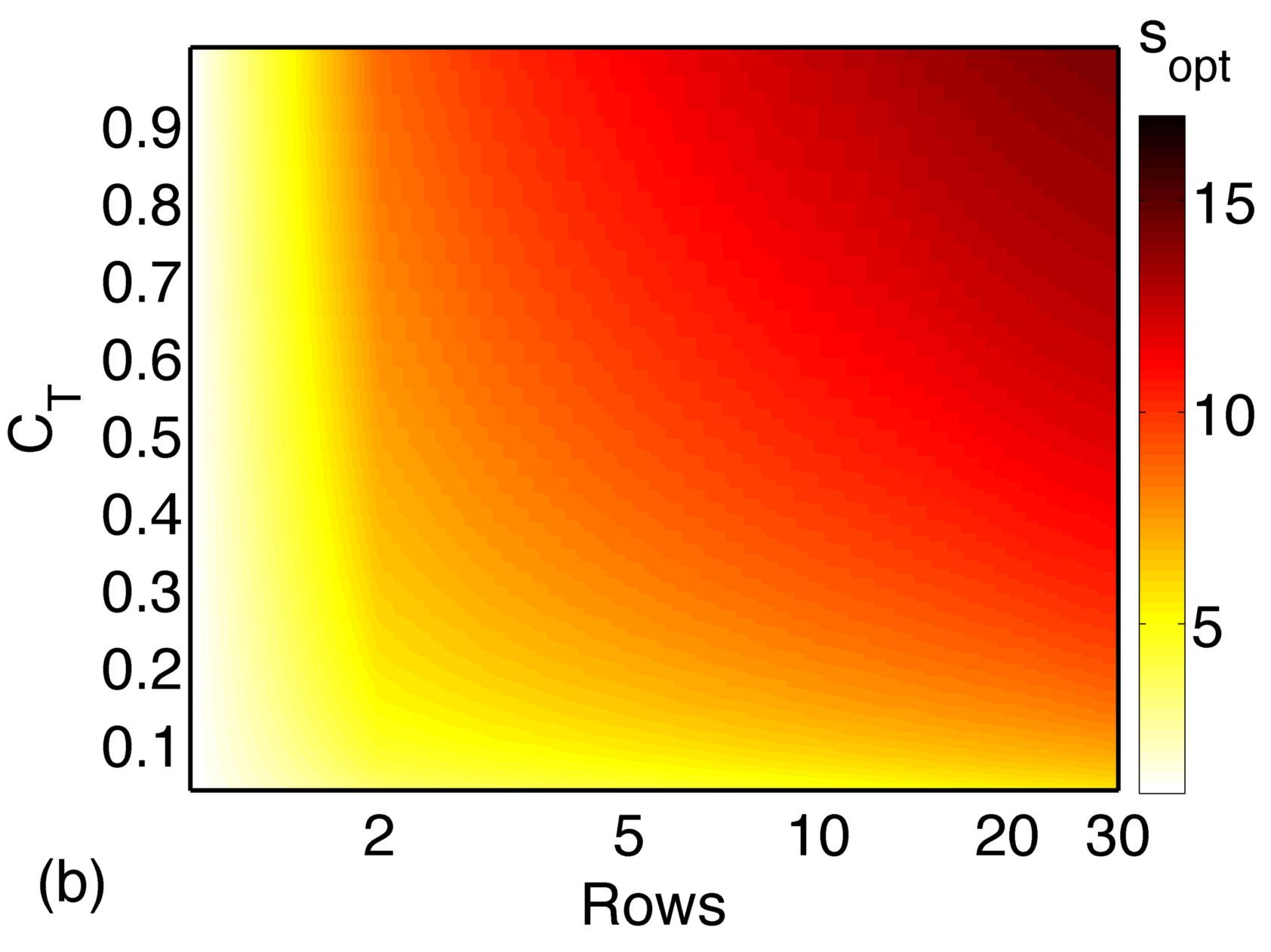}}
\caption{Optimal geometrical mean turbine spacing for (a) a wind farm with $6$ rows as function of the dimensionless cost ratio $\alpha$ and the turbine thrust coefficient $C_\mathrm{T}$ and (b) for $\alpha=2000$ as function of the number of turbine rows and the turbine thrust coefficient $C_\mathrm{T}$.}
\label{figure8}
\end{center}
\end{figure}

Figure \ref{figure8} reveals the effect of the turbine thrust coefficient $C_\mathrm{T}$ on the optimal turbine spacing as function of the wind farm length and the dimensionless cost ratio $\alpha$. Figure \ref{figure8}a shows that for a wind farm with $6$ downstream turbine rows the optimal turbine spacing becomes larger when the turbine thrust coefficient $C_\mathrm{T}$ or the cost ratio $\alpha$ is increased. This effect is most pronounced for high $C_\mathrm{T}$ coefficients, i.e.\ when the turbines produce more power and generate relatively strong wakes. For very low $C_\mathrm{T}$ values the optimal turbine spacing is significantly smaller as the generated wakes are relatively weak. Here it must be emphasized that operating on a higher $C_\mathrm{T}$ coefficient is beneficial as then more power is produced. Therefore the upper part of figure \ref{figure8} is most relevant as this shows the region of $C_\mathrm{T}$ values found in normal operational wind farms ($C_\mathrm{T}\approx0.75$) \cite{por13}. The low $C_\mathrm{T}$ range would only be relevant if all turbines would operate at these low $C_\mathrm{T}$ values, which is in general not the case. Here it is noted that the data presented in the figure assume that all turbines in the wind farm operate in the region II \cite{joh04} in which the $C_\mathrm{T}$ value is constant as function of the wind speed. The constant thrust coefficient in this regime is a result of the blade rotation frequency that is adjusted appropriately to the incoming wind. As is indicated in the paper by Port\'e Agel {\it et al.} \cite{por13} the turbines operate most of the time in this regime and therefore this regime is particularly interesting. For very strong winds it can happen that the $C_\mathrm{T}$ value for some turbines, particularly the ones on the first row, is limited by feathering the blades. As the aim of the paper is to reveal the general trends this is not considered here as the effect will depend on the wind speed distribution. From figure \ref{figure8} we can see that lowering of the average $C_\mathrm{T}$ coefficient results in a smaller effective optimal spacing, however for small variations of the $C_\mathrm{T}$ value (of the order of 0.1) the influence on the optimal spacing is relatively modest. Figure \ref{figure8}b shows the optimal spacing as a function of the average $C_\mathrm{T}$ and the length of the wind farm for a fixed cost ratio $\alpha=2000$. In agreement with figure \ref{figure6} we see that optimal spacing increases with increasing wind farm size. \\

\section{Discussion and conclusion} \label{Section_conclusion}
In the study presented here we have used a physics based model that parametrizes the effective roughness height of wind farms and explored the implications on the optimal spacing among the wind turbines. Earlier work by Meyers and Meneveau \cite{mey12} using this approach focused on the optimal spacing in infinitely large wind farms and found that for realistic cost ratios of the turbine and land costs the optimal spacing may be considerable larger than used in existing wind farm designs. In an effort to explain this apparent discrepancy we have now incorporated the effect of the wind farm length. The model presented here shows that the optimal turbine spacing depends strongly on wind farm length, i.e.\ the number of rows in the downstream direction, in such a way that the predicted optimum is consistent with spacings found in operational wind farms of small or moderate size. For much larger farms, the larger optimal spacing found for infinite wind farms by Meyers and Meneveau \cite{mey12} is confirmed.

Obviously, the conclusions reached here are subject to considerable limitations. First of all the current parametrization makes no distinction between the spanwise and streamwise spacing between the turbines or their relative placing with respect to the prevailing wind direction. A comparison of the model with LES results and measurements in Horns Rev and Nysted shows that the model predictions are consistent with observations for staggered wind farms.  In the LES the flow is essentially driven by a pressure gradient. As is explained in the appendix of Ref.\ \cite{cal10} this is equivalent to assuming a given value for the transverse geostrophic wind and this approach is expected to give meaningful results as the flow in the inner region (up to $0.15-0.20H$) is not influenced by external effects such as Coriolis forces. Finally, we note that the parametrization of the interaction between the wind farm and the atmospheric boundary layer is derived for neutral stratification conditions and assumes a flat terrain with no topography effects. In many land based wind farms the topography will influence the local interactions and thus the optimal siting of the turbines. Also, for large offshore wind farms it may be difficult to specify the `per turbine cost' and the `per area cost', and can depend on the connectivity cost, distance to the coast, typical sea state, etc. In addition, the turbine spacings also influence the unsteady structural loadings and therefore the maintenance costs, which are not taken into account here. In short, the shown results provide a guideline for the spacing that should be used in wind farms to get the highest power per unit cost and how this spacing depends on some of the main design parameters of a wind farm. However, it is emphasized that for the design of an actual wind farm local effects and restrictions should always been taken into account.

\ack{The author thanks Charles Meneveau and Dennice Gayme for valuable conversations and comments. This work is funded in part by the research program `Fellowships for Young Energy Scientists' (YES!) of the Foundation for Fundamental Research on Matter (FOM) supported by the Netherlands Organization for Scientific Research (NWO), and by the National Science Foundation grant number 1243482 (the WINDINSPIRE project). This work used the Extreme Science and Engineering Discovery Environment (XSEDE), which is supported by the National Science Foundation grant number OCI-1053575 and the LISA cluster of SURFsara in the Netherlands.}


\FloatBarrier

\section*{Appendix: Horns Rev and Nysted layouts}
Horns Rev consists of 80 Vestas 2 MW turbines each with a hub-height of $z_\mathrm{h}=70$m and a rotor diameter $D=80$m with a streamwise turbine spacing $s_\mathrm{x}=7.00$ and a spanwise turbine spacing $s_\mathrm{y} = 6.95$ as the layout parameters for the aligned configuration (270$^\circ$). The main data available from Horns Rev span the wind direction range 255$^\circ$ to 285$^\circ$ and the 285$^\circ$ wind direction is closest to what can be considered as a staggered configuration \cite{bar09c,bar11}. The Nysted wind farm consists of 72 Siemens 2.3MW turbines with a hub-height $z_\mathrm{h}=69$m and a rotor diameter $D=82.4$m.  The aligned configuration of Nysted at 278$^\circ$ has a streamwise spacing $s_\mathrm{x}=10.3$ and a spanwise spacing $s_\mathrm{y}=5.8$. For Nysted the main available data span the wind direction range 263$^\circ$ to 293$^\circ$ and the 293$^\circ$ wind direction is closest to what can be considered as a staggered configuration \cite{bar09c,bar11}.

\begin{figure}[!ht]
\begin{center}
\subfigure{\includegraphics[width=0.47\textwidth]{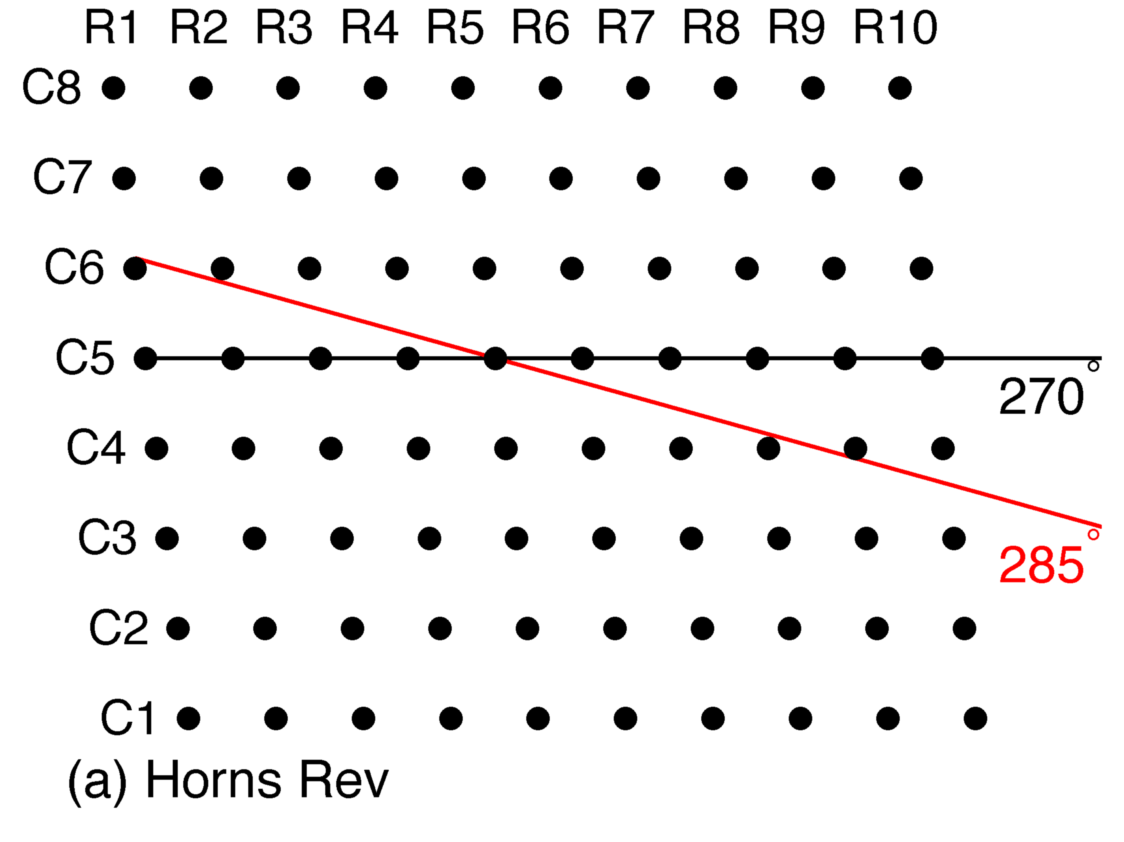}}
\subfigure{\includegraphics[width=0.47\textwidth]{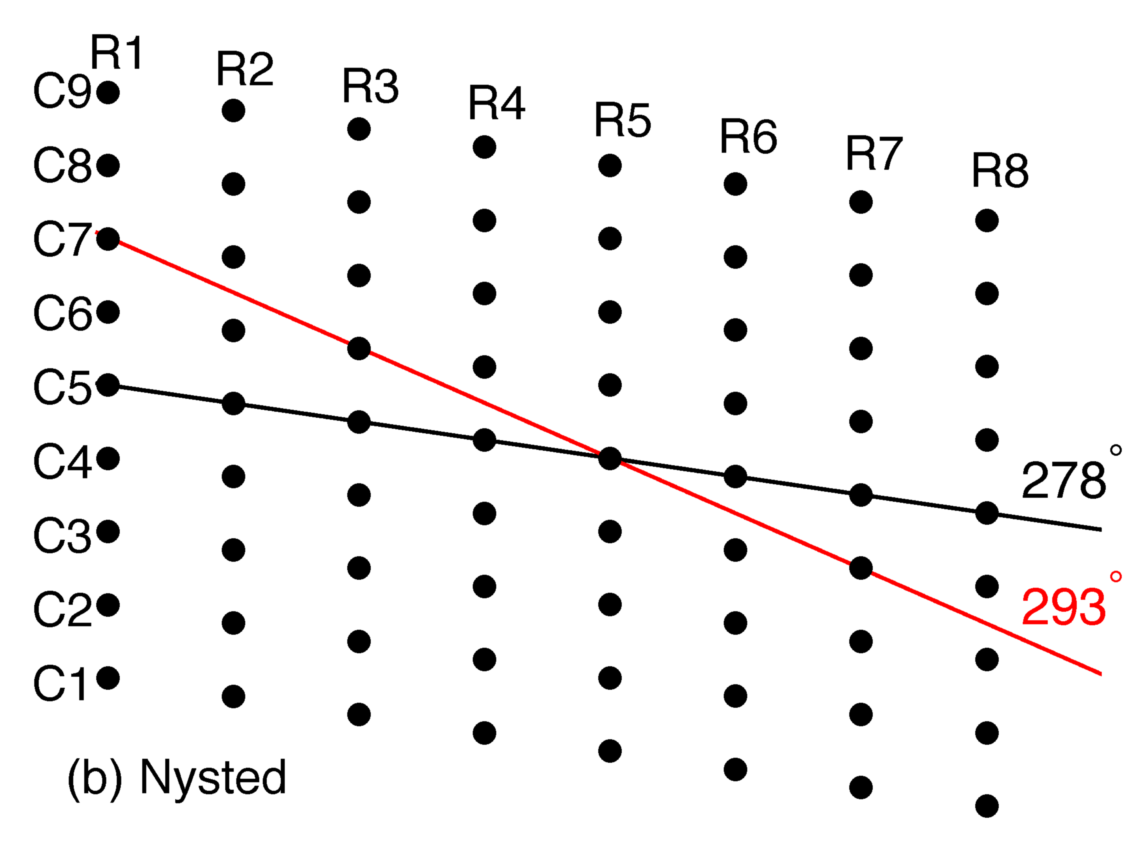}}
\caption{Layout of the (a) Horns Rev and (b) Nysted wind farms. The column and row numbers are also indicated.}
\label{figure9}
\end{center}
\end{figure}

\end{document}